\documentclass[aps,prl,twocolumn,showpacs,floats,preprintnumbers,amsmath,amssymb]{revtex4}

\usepackage{graphicx}
\usepackage{dcolumn}
\usepackage{bm}

\usepackage[pdffitwindow=false,bookmarks=true,colorlinks=true,citecolor=blue,linkcolor=blue]{hyperref}

\begin{document}

\preprint{(submitted to Phys. Rev. B)}

\title{Evaluation of endohedral doping of hydrogenated Si fullerenes as a route to magnetic Si building blocks}

\author{Dennis Palagin}
\affiliation{Department Chemie, Technische Universit{\"a}t M{\"u}nchen, Lichtenbergstr. 4, D-85747 Garching, Germany}

\author{Karsten Reuter}
\affiliation{Department Chemie, Technische Universit{\"a}t M{\"u}nchen, Lichtenbergstr. 4, D-85747 Garching, Germany}

\date{\today}

\begin{abstract}
Density-functional theory based global geometry optimization is used to scrutinize the possibility of endohedral doping of hydrogenated Si fullerenes as a route to Si nanostructures with high magnetic moments. In contrast to previous suggestions, our unbiased sampling finds the smallest Si$_{16}$H$_{16}$ endohedral cage generally too small to encapsulate $3d$ metal dopant atoms. For the next larger fullerene-like cage though, we identify perfectly symmetric $M$@Si$_{20}$H$_{20}$ ($M$=Co, Ti, V, Cr) cage structures as ground states. These structures conserve the high spin moment of the dopant atom and therewith underscore the potential of this Si nanoform for novel cluster-based materials with unique magnetic properties.
\end{abstract}


\maketitle

The prevalence of silicon in semiconductor industry has raised great interest in silicon-based nanostructures that could act as building blocks for new easy-to-integrate and engineered materials. In contrast to the compact and reactive form of pure Si clusters, endohedral metal doping was found to produce saturated fullerene and other polyhedral cage structures, and therewith an intriguing novel nanoform of Si with great promise e.g. for optoelectronic device applications \cite{jackson96,sun02,hiura01,kumar01,kumar02,kumar04,kumar06,khanna05,khanna06,nakajima07,torres07,torres11,gramzow10,fielicke11}. In these structures the cage geometry is stabilized through a strong interaction with the encapsulated dopant atom, which unfortunately goes hand in hand with a quenching of the dopant spin moment \cite{palagin11}. The recent suggestion that also hydrogen termination of Si clusters could yield an (empty) fullerene configuration \cite{kumar03,kumar07,kumar09} has thus raised hopes that metal-doping corresponding hydrogenated fullerenes would yield cage structures with minimized $M$-Si interaction \cite{kumar09}. With the atomic character of e.g. magnetic dopants then likely conserved, this would offer a route to specifically develop Si fullerene species with large magnetic moments for magneto-electronic applications.

In order to further explore and substantiate this idea we perform density-functional theory (DFT) based global geometry optimizations of the smallest hydrogenated endohedral cage, Si$_{16}$H$_{16}$, for a range of metal dopants. In order to address a possible influence of the size of the dopant atoms, this range comprises not only previously discussed high magnetic moment species like Ti or Cr \cite{kumar09}, but extends over the entire $3d$ series. While our extensive unbiased sampling indeed confirms the cage-like geometry as global ground state of empty Si$_{16}$H$_{16}$, this is unfortunately not the case for any of the $3d$ dopants. Instead of a symmetric metal encapsulation, strongly distorted or even broken cages are significantly more stable in all cases. Moreover, the stronger $M$-Si interaction in these ground-state structures leads to the same quenching of the magnetic moment as in the regular $M$@Si$_{16}$ fullerenes. Attributing these discouraging findings to the insufficient space inside Si$_{16}$H$_{16}$, we therefore extend our study to the next larger fullerene-like structure, Si$_{20}$H$_{20}$ \cite{kumar03}. Testing the high magnetic moment sequence Ti, V, Cr, and Co we now obtain in all cases perfectly symmetric endohedral cage geometries as ground states. Spin density distribution analysis furthermore reveals that the high spin states of these structures come indeed primarily from conservation of the dopant magnetic moment, i.e. the tested sequence already offers an intriguing toolbox of Si nanoforms with septet to quartet spin, respectively. Confirming also the next larger Si$_{24}$H$_{24}$, Si$_{26}$H$_{26}$ and Si$_{28}$H$_{28}$ cages in the fullerene sequence as corresponding ground-state geometries, this strongly supports the idea of generating promising magnetic building blocks by doping sufficiently-sized hydrogenated Si cages.

All ground-state total energy calculations in this work have been performed with the all-electron full-potential DFT code FHI-aims \cite{blum09}, treating electronic exchange and correlation within the generalized-gradient approximation functional due to Perdew, Burke and Ernzerhof (PBE) \cite{perdew96}. All sampling calculations were done with the "tier2" atom-centered basis set and using "tight" settings for the numerical integrations. Stability of the identified minima has been confirmed by frequency analysis. For the ensuing electronic structure analysis of the optimized geometries, the electron density was recomputed with an enlarged "tier3" basis set \cite{blum09}. As in our previous work on metal-doped Si clusters \cite{gramzow10,palagin11} systematic convergence tests indicate that these settings are fully converged with respect to the target quantities, here the energetic difference of isomers, as well as total electron and spin density distributions. For the actual global geometry optimization procedure we relied on the basin-hopping (BH) algorithm \cite{doye97,wales00}, which samples the huge configurational space through consecutive jumps from one local minimum of the potential energy surface to another. In our specific BH implementation \cite{gehrke09,gramzow10,palagin11} we achieve this by randomly moving one or all atoms in the cluster to a new position within a sphere of 1.5\,{\AA}, each time followed by a local geometry optimization and acceptance or rejection of the thus created structure on the basis of a Metropolis algorithm.

\begin{figure*}
\includegraphics[width=\linewidth]{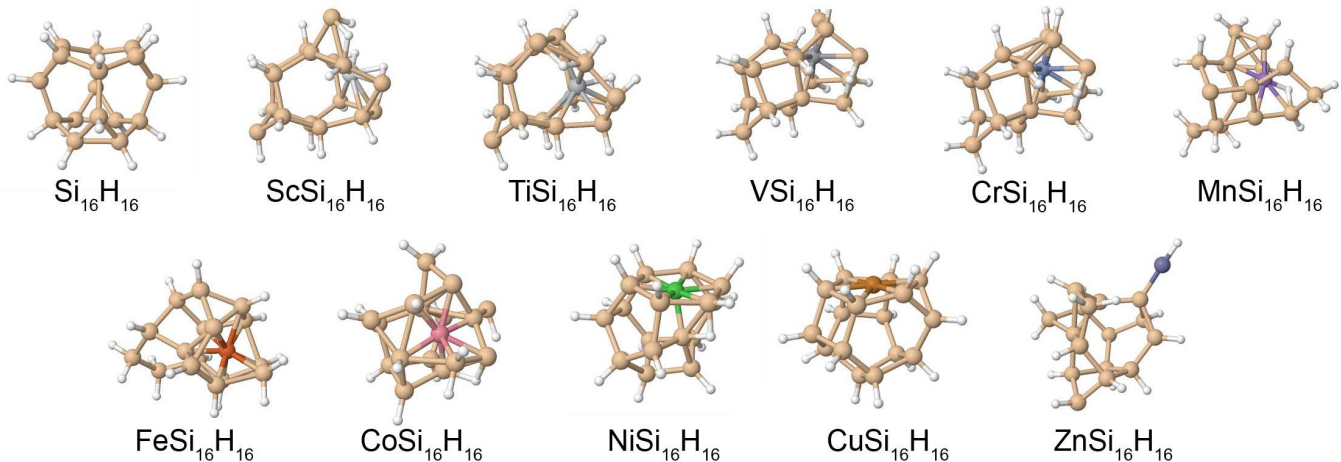}
\caption{Identified ground-state geometry of Si$_{16}$H$_{16}$, as well as of $M$@Si$_{16}$H$_{16}$ clusters (with $M$ covering the $3d$ metal series).}
\label{fig1}
\end{figure*}

The starting point of our investigation is an extended configurational search, which even after thousands of trial structures confirms that the ground-state structure of Si$_{16}$H$_{16}$ is indeed the perfectly symmetrical tetrahedral (T$_{d}$) cage shown in Fig. \ref{fig1} \cite{kumar03}. Featuring Si atoms at distances of 2.95\,{\AA} and 3.17\,{\AA} from the center, this cage provides in principle enough geometric space to host an endohedral dopant atom, with the hydrogenation hoped to minimize the actual $M$-Si interaction. Unfortunately, the global geometry optimization results for $M$Si$_{16}$H$_{16}$ clusters with metal dopant atoms ranging over the entire $3d$ series from Sc to Zn reveal a different picture. In all cases only a few BH steps are required to identify significantly more stable geometries, which correspond to severely distorted cages or more compact structures that push the dopant atom from the center to the cluster fringe (or even expel it completely in case of Zn), cf. Fig. \ref{fig1} \cite{supplement}. With the exception of Zn (0.37\,eV) the energetic difference of these new structures to the symmetric endohedral cage is of the order of 1\,eV or larger, which makes a kinetic trapping in the high-energy tetrahedral cage geometry rather unlikely. It also suggests that the obtained energetic order is not affected by the well-known limitations of the employed semi-local DFT functional. To further validate this, we nevertheless evaluated the energetic differences also with the hybrid PBE0 functional \cite{pbe0}, without obtaining any qualitative changes.

As already indicated by the typically high Si coordination, the identified low-energy isomers exhibit a rather strong $M$-Si interaction involving substantial charge rearrangements. Not surprisingly, this goes along with an almost complete quench of the dopant spin moment. Clusters having an even number of electrons are singlets (Ti, Cr, Fe, Ni and Zn doping), clusters with odd number of electrons are doublets (Sc, V, Mn, Co and Cu doping), with again no changes observed when repeating the calculations at the PBE0 hybrid functional level. In case of the doublets, the unpaired electron is hereby not even necessarily centered on the dopant atom: For both Co and Sc the spin density is either partly or even mainly located on Si atoms, respectively. Obviously, this completely rules out the original proposition of using these nanomaterials as high magnetic moment Si building blocks.

\begin{figure}
\includegraphics[width=\linewidth]{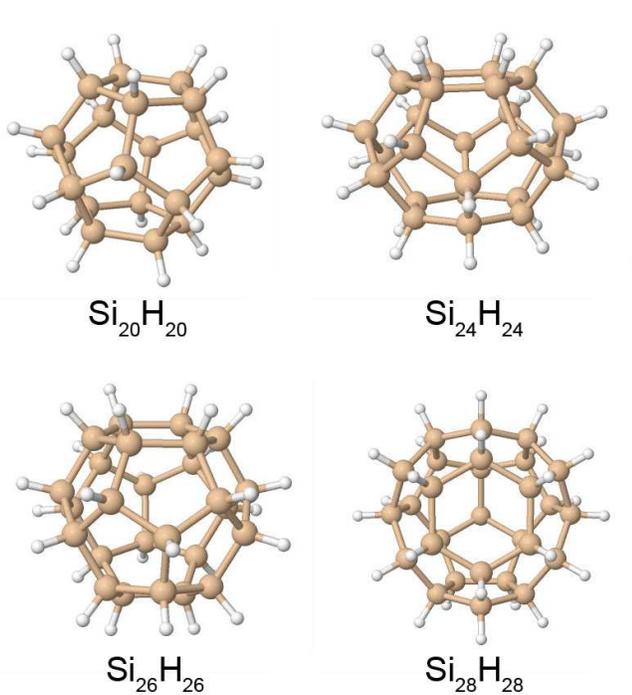}
\caption{Identified hollow cage ground-state geometries of the next larger hydrogenated Si fullerenes.}
\label{fig2}
\end{figure}

Suspecting an insufficient space inside the cage as reason for these findings we proceed to the next larger hydrogenated fullerene-like cage Si$_{20}$H$_{20}$, for which extensive sampling again confirms the empty cage geometry shown in Fig. \ref{fig2} as ground-state structure. Intriguingly, the increased radius of 3.33\,{\AA} of this icosahedral ($I_h$) cage now seems to be large enough to even accommodate most space-demanding high spin-state dopant atoms. Focusing on the sequence Co, Ti, V, and Cr with quartet, quintet, sextet and septet atomic spin moments, respectively, the endohedrally-doped symmetric cage corresponds each time to the most stable structure found in our extensive sampling runs. Moreover, the energetic gap to the second lowest identified isomer, which then corresponds to a distorted or broken cage structure, is in all cases larger than ~0.5\,eV. From this we again do not suspect our findings to be jeopardized by the limitations of the employed semi-local DFT functional; an interpretation that we find confirmed by hybrid PBE0-level calculations that yield for all four dopant atoms the same energetic order of ground-state cage and second lowest isomer as in the PBE case.

For this fullerene size we thus find a beautiful confirmation of the hydrogenation concept, considering that previous global geometry optimization work clearly identified the corresponding endohedral $M$@Si$_{20}$ cages ($M=$Ti, V, Cr) as metastable, with the ground-states of these unhydrogenated clusters instead given by heavily distorted and reactive structures \cite{gramzow10}. In contrast to the again almost complete spin quench in the latter structures, the doped hydrogenated fullerenes furthermore perfectly conserve the high magnetic moments of their metal dopants, i.e. Co@Si$_{20}$H$_{20}$, Ti@Si$_{20}$H$_{20}$, V@Si$_{20}$H$_{20}$, Cr@Si$_{20}$H$_{20}$ exhibit quartet, quintet, sextet and septet spin moments, respectively. Analysis of the spin density distribution confirms that the unpaired electrons indeed reside predominantly on the central metal atom, which is in line with the obtained almost negligible hybridization of the occupied metal $3d$ states. The metal-cage interaction instead seems largely mediated via the metal $4s$ state, which rationalizes the stabilization of a symmetric dopant position in the center of the cage. 

\begin{figure}
\includegraphics[width=\linewidth]{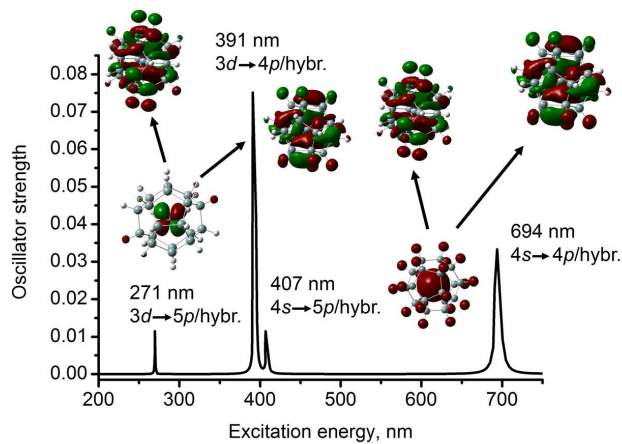}
\caption{Calculated TD-DFT optical excitation spectrum of the endohedral Cr@Si$_{20}$H$_{20}$ fullerene ground-state, with the insets illustrating the dominant Kohn-Sham orbitals behind the individual transitions.}
\label{fig3}
\end{figure}

Despite considerable binding energies, which at the PBE-level exceed 1\,eV \cite{supplement} and therewith stabilize the endohedral cage geometry, the atomic character of the metal dopant atom is thus largely conserved. This is also apparent in the optical excitation spectra, computed within the time-dependent DFT (TD-DFT) linear response formalism as implemented in the Gaussian03 suite \cite{gaussian03,tddft_details}. Figure \ref{fig3} illustrates this for the Cr@Si$_{20}$H$_{20}$ case. In stark contrast to corresponding spectra of e.g. the broken cage structures of $M$Si$_{16}$H$_{16}$, only a few well-defined transitions are apparent in the optical range. These transitions are energetically shifted and changed in their order in comparison to the corresponding spectra of the isolated atomic dopants. Nevertheless the "atomic" character is preserved, i.e. apart from their magnetic properties the endohedrally-doped hydrogenated fullerenes also exhibit promising optical properties.

In summary, we have thus systematically scrutinized the possibility of stabilizing endohedrally doped Si fullerenes through hydrogenation. Our unbiased global geometry optimization based on semi-local and hybrid functional DFT undoubtedly revealed that the smallest hydrogenated Si$_{16}$H$_{16}$ cage focused on in previous work is generally too small to encapsulate $3d$ transition metals. On a conceptual level this underscores the danger of basing nanoscale materials design through predictive-quality theory on incomplete explorations of the vast configurational space, like simple comparisons of chemically intuitive candidate structures. For the next larger fullerene-like cage though, our first-principles sampling indeed identifies perfectly symmetric $M$@Si$_{20}$H$_{20}$ ($M$=Co, Ti, V, Cr) cage structures as ground states. Compared to distorted or broken cage geometries, these structures are stabilized through substantial delocalized $4s$-cage interaction, while nevertheless largely retaining the atomic character of the metal dopant. With respect to magnetic properties, the confirmed quartet, quintet, sextet and septet spin moments of Co@Si$_{20}$H$_{20}$, Ti@Si$_{20}$H$_{20}$, V@Si$_{20}$H$_{20}$, and Cr@Si$_{20}$H$_{20}$, respectively, thus already offer a nice toolbox of unreactive building blocks with high magnetic moments. Nicely shielded by the fullerene cage, the encapsulated dopants furthermore also conserve atom-like optical properties. This strongly suggests endohedral doping of hydrogenated fullerenes as a viable route to novel cluster-based materials for magneto-optic applications. This vision gets further support by extensive sampling runs that also suggest empty cage geometries as ground-states for the next larger hydrogenated Si fullerenes up to Si$_{28}$H$_{28}$, cf. Fig. \ref{fig2}. Eventually, these cages will offer enough space to also host multi-core dopants. Encapsulation into hydrogenated Si fullerenes thus appears as a promising avenue to generally protect metal clusters from reactive environments and make their unique material's properties available in applications.

Funding within the DFG Research Unit FOR1282 and support of the TUM Faculty Graduate Center Chemistry is gratefully acknowledged.


\begin{thebibliography}{29}
\bibitem[{tdd()}]{tddft_details}
\bibinfo{note}{To calculate the excitation spectrum of the cluster the first
  100 transitions have been calculated using PBE0 functional and LANL2DZ
  \cite{lanl2dz} basis set within the TD-DFT \cite{tddft_paper} approach.}
\end{thebibliography}
\end{document}